\documentclass[aps,physrev,twocolumn,superscriptaddress]{revtex4-2}
\usepackage{graphicx}
\usepackage{longtable}
\usepackage{amsmath}
\usepackage[overload]{textcase} 

\begin{document}

% \preprint{}

\title{Charge density wave in intermetallic oxides R$_5$Pb$_3$O (R = La and Ce)}

\author{Rafaela F. S. Penacchio}
\affiliation{Ames National Laboratory, US DOE, Iowa State University, Ames, IA, USA}
\affiliation{Department of Physics, Iowa State University, Ames, Iowa 50011, USA}
\affiliation{Institute of Physics, University of S{\~{a}}o Paulo, S{\~{a}}o Paulo, SP, Brazil}

\author{Siham Mohamed}
\affiliation{Ames National Laboratory, US DOE, Iowa State University, Ames, IA, USA}
\affiliation{Department of Chemistry, Iowa State University, Ames, Iowa 50011, USA}

\author{Haley A. Harms}
\affiliation{Ames National Laboratory, US DOE, Iowa State University, Ames, IA, USA}
\affiliation{Department of Physics, University of Northern Iowa, Cedar Falls, IA, 50614 USA}
% \affiliation{Department of Chemistry, Iowa State University, Ames, Iowa 50011, USA}

\author{Lin-Lin Wang}
\affiliation{Ames National Laboratory, US DOE, Iowa State University, Ames, IA, USA}

\author{Sergey L. Bud’ko}
\affiliation{Ames National Laboratory, US DOE, Iowa State University, Ames, IA, USA}
\affiliation{Department of Physics, Iowa State University, Ames, Iowa 50011, USA}

\author{Paul C. Canfield}
\affiliation{Ames National Laboratory, US DOE, Iowa State University, Ames, IA, USA}
\affiliation{Department of Physics, Iowa State University, Ames, Iowa 50011, USA}

\author{Tyler J. Slade}
\affiliation{Ames National Laboratory, US DOE, Iowa State University, Ames, IA, USA}

\date{\today}

\begin{abstract}

The R$_5$Pb$_3$O family was discovered decades ago, but has remained largely unexplored. Here, we report single crystal growth and basic characterization for the La and Ce members of this family. At room temperature, these compounds adopt a tetragonal structure ($I4/mcm)$, where R and Pb atoms form linear chains along the $c$-axis. We identify a second-order structural phase transition at 260 K and 145 K for R = La and Ce, respectively. Single crystal X-ray diffraction reveals a lattice modulation below the transition temperature, resulting in R-Pb pairs in the $z$ direction. The broken symmetry in the low-temperature phases results in a primitive structure with space group $P4/ncc$. Transport and diffraction measurements, in agreement with density functional theory calculations, support that the R$_5$Pb$_3$O (R = La and Ce) series hosts an electron-phonon coupling driven charge density wave (CDW) at low temperatures. The CDW ordering temperature is suppressed by more than 100 K by the La to Ce substitution, suggesting high pressure-sensitivity. Therefore, this family offers the potential for investigating competing orders in oxides, with heavier rare-earth members still to be explored.

\end{abstract}

\maketitle

\section{Introduction}

Charge density waves (CDWs) are structural phase transitions that occur in metals when an instability at the Fermi surface drives a redistribution of the electronic density \cite{peierls1955quantum,frohlich1954theory,thorne1996}. This phase transition is associated with a modulation of the conduction electron density and a corresponding gaping (or partial gaping) of electronic bands near the Fermi level. Owing to finite electron-phonon coupling, the electronic modulation causes the nuclei to assume new equilibrium positions, so that the electronic instability is accompanied by a simultaneous structural distortion \cite{PhysRevLett.51.138,woll1962}. Whereas CDWs are well understood theoretically and experimentally in one-dimensional systems, both the underlying mechanism and characteristic behavior of CDWs remain controversial in two- and three-dimensional metals \cite{johannes2008fermi,zhu2017misconceptions,PhysRevB.73.205102}. Furthermore, CDWs are often found in competition with other ordered states, frequently including superconductivity \cite{morosan2006superconductivity,chang2012direct,ortiz2021superconductivity,akiba2022observation}. For these reasons, identifying and studying new model systems with density wave transitions is of general interest.

The R$_5$Pb$_3$O family of materials crystallizes in an interstitial derivative of the tetragonal Cr$_5$B$_3$ structure, in contrast to many hexagonal R$_5$Pb$_3$Z compounds (Z = B, C, P, Si, Ge, Sb) \cite{guloy1994}. The crystal structure of these tetragonal phases contains linear chains of alternating R and Pb atoms that extend along the $c$-axis. Single crystals of R = La and Ce members were grown decades ago and, interestingly, the La/Ce site was found to exhibit anomalously large displacement parameters in the $z$ direction \cite{guloy1992, macaluso2004}. At that time, this anomaly was interpreted as a signature of static disorder related to the random formation of R-Pb pairs. Despite the crystal structure being known for decades, limited information has been published concerning the physical properties of these materials. Macaluso et al. found that Ce$_5$Pb$_3$O orders ferrimagnetically at 46 K \cite{macaluso2004}; however, to our best knowledge, this is the only study of physical properties reported for the R$_5$Pb$_3$O series. 

Here, we explore the crystal structure and physical properties of R$_5$Pb$_3$O single crystals. Transport measurements reveal a CDW-like transition at 260\,K and 145\,K for R = La and Ce, respectively. Temperature-dependent single crystal X-ray diffraction shows that symmetry is broken in the R/Pb chain in the low-temperature phase, resulting in R-Pb pairs, similar to the dimerization in Peierls' picture. Density functional theory calculations indicate that the CDW in these compounds is driven primarily by strong electron-phonon coupling, instead of Fermi surface nesting. In the high-temperature phase, the Fermi level lies in a region of high density of states, and the transition is driven by the softening of an optical phonon at the $Z$ point. The calculated distortion and corresponding low-temperature structure agree well with the experimental data. Interestingly, we also find that the thermal displacement parameters of R atoms in the chains are anomalously ellipsoidal, with $U_{33}/U_{11}$ maximized at the structural transition temperature. We interpret this behavior as an experimental signature of the underlying instability and corresponding phonon softening that drives the transition. The 100 K difference in ordering temperatures between La and Ce phases suggests that the transition may be very sensitive to external perturbations, such as applied pressure, stress, or chemical substitution; therefore, this work established that the R$_5$Pb$_3$O series is a potential platform to explore competing orders, with other members of this family still to be investigated.

\section{Experimental details}

\subsection{Single crystal growth}

Single crystals of La$_5$Pb$_3$O were obtained by the high-temperature solution growth method \cite{canfield2019}. Our original intent was to find new intermetallic compounds containing La, Ni, and Pb. In the first attempts, elemental La (Ames Laboratory, 99.9+\%), Ni pieces (American Elements, 99.99\%), and Pb (COMNICO Electronic Metals, 99.999\%) were weighted in molar ratios of La$_{45}$Ni$_{50}$Pb$_5$, and loaded into a Ta crucible set with home-made Ta caps and filter \cite{canfield2001, canfield2016}. The Ta crucibles were first sealed under an Ar atmosphere using an arc melter, and then flame-sealed into fused silica ampules that were backfilled with $\approx 1/6$ atm Ar gas. Using a box furnace, the ampules were warmed to 1150\,$^\circ$C over $\approx$ 6 h and held at that temperature for 6 h. Then, the furnace was gradually cooled to 750\,$^\circ$C over 70 -- 80 h. Upon reaching the desired temperature, the excess liquid phase was decanted in a centrifuge with metal cups and rotors \cite{canfield2019}. After cooling to room temperature, the tubes were opened to reveal a few rod-like crystals with typical dimensions of $\approx 1$ mm, as shown in the inset of Fig. \ref{fig:diffraction}(c). Using powder and single crystal X-ray diffraction (described in the following sections), we identified the rod-like crystals as the suboxide La$_5$Pb$_3$O \cite{guloy1992}.

The growth of an oxide, La$_5$Pb$_3$O, from a nominally oxygen-free melt was surprising to us. Interestingly, previous literature on R$_5$Pb$_3$O phases also reported the growth of these compounds as subproducts of R-T-Pb reactions (T = transition metals). For example, Guloy and Corbett identified La$_5$Pb$_3$O as an impurity in La-Mn-Pb reactions \cite{guloy1994}, Macaluso et al. produced Ce$_5$Pb$_3$O single crystals when searching for new ternary compounds in the Ce-Co-Pb phase space \cite{macaluso2004}, and Yan grew R$_5$Pb$_3$O (R = La, Ce, Pr, Nd, and Sm) single crystals using melts with nominal compositions of R$_7$Co$_2$Pb \cite{yan2015}. Yan speculated that the oxygen was introduced into the melt by partial reaction between R elements and the Al$_2$O$_3$ crucible used to contain their reaction. In our reactions, because Ta crucibles (inert to reactions with rare-earth elements) were used as containers, we conclude that the oxygen must have come from the surfaces of one or more of the starting elements, most likely La. In support of this hypothesis, we found that subsequent attempts to reproduce the initial growth gave inconsistent results, with reactions yielding La$_5$NiPb$_3$, La$_5$Pb$_3$, and/or La$_5$Pb$_3$O phases in varying quantities from batch to batch. These results indicated that a small quantity of surface oxide (inherently an unquantifiable entity) was being introduced in an unpredictable amount into our growths. 

To moderately control the amount of oxygen in the growth, we added up to 3.5 mol \%, relative to the starting quantity of La, of La$_2$O$_3$ (Ames National Laboratory, 5-9's) into an initial La$_{45}$Ni$_{50}$Pb$_5$ mixture. The growths were contained in Ta crucibles and subject to the same heat treatment described above, and we found that these attempts reproducibly resulted in La$_5$Pb$_3$O single crystals. We also succeeded in growing La$_5$Pb$_3$O by adding 2.5\% (relative to the starting quantity of Pb) of PbO (Morton Thiokol, 99.9\%) into the starting melt. Powder X-ray diffraction patterns on the products of different growths are shown in Fig. \ref{fig:powderLaAttempts} in the supporting information (SI). Our attempts to produce La$_5$Pb$_3$O crystals by adding La$_2$O$_3$ to La-Pb melts were not successful, implying that, despite not being incorporated into the product, the transition metal likely plays an important role in the melt chemistry. Given that La-Ni mixtures are low melting ($<800^\circ$C) at La-rich compositions ($\approx$ 50-80 $\%$ La), we speculate that the La- and Pb-oxides are partially soluble in La-Ni melts. Compared to relying on surface oxidation or reaction with the crucible to provide oxygen, our results show that controlled addition of La- or Pb-oxides into La-Ni-Pb melts favors the growth of the suboxide La$_5$Pb$_3$O, whereas oxygen-free, or deficient, melts result in the La$_5$NiPb$_3$ and La$_5$Pb$_3$ phases.

Single crystals of Ce$_5$Pb$_3$O were grown by adding up to 2.5\% (relative to the starting quantity of Pb) of PbO (Morton Thiokol, 99.9\%) into an initial Ce$_{45}$Ni$_{50}$Pb$_5$ mixture contained in a Ta crucible, where the growth procedure and heat treatment were otherwise the same as described above for La$_5$Pb$_3$O. Powder X-ray diffraction, shown in Fig. \ref{fig:powderZX711} in the SI, indicated Ce$_5$Pb$_3$O as the main phase. We found that unlike La$_5$Pb$_3$O, which formed rod-like crystals, the Ce$_5$Pb$_3$O formed either rods or 3D-block-like crystals. Additionally, the Ce-containing growths normally yielded mixtures of Ce$_5$Pb$_3$O and hexagonal Ce$_5$Pb$_3$. Prior to subsequent measurements, rod-like crystals of Ce$_5$Pb$_3$O were isolated and the phase confirmed with single crystal x-ray diffraction.

We also succeeded in growing La$_5$Pb$_3$O and Ce$_5$Pb$_3$O by following the protocol outlined by Yan and Macaluso et al. \cite{yan2015, macaluso2004}, in which the starting melt has composition R$_7$Co$_2$Pb and is contained in an Al$_2$O$_3$ crucible. However, elemental analysis shows a small, but observable, Al contamination in the R$_{5}$Pb$_{3}$O crystals obtained from this method, see Fig. \ref{fig:EDSRT} in the SI. Therefore, the following discussion is based on single crystals grown in Ta crucibles. Comparisons to Al$_2$O$_3$-grown samples are provided in the SI. As described in previous literature, La$_5$Pb$_3$O and Ce$_5$Pb$_3$O crystals were found to be very oxygen- and/or moisture-sensitive. The bright surface of the as-grown samples darkened after minutes of air exposure, with samples decomposing within a few days. Therefore, the samples were stored in an N$_2$ glovebox until needed for characterization. 

\subsection{X-ray diffraction}

Single crystal X-ray diffraction was performed using a Rigaku XtaLab Synergy-S diffractometer with Ag radiation ($\lambda = 0.56087$ \AA), in transmission mode, operating at 65 kV and 0.67 mA. The samples were held in a nylon loop with Apiezon N grease, and the temperature was controlled using an Oxford Cryostream 1000. At least 20 minutes were waited at each temperature to ensure thermal stabilization before starting the measurement. The total number of collected runs and images was based on the strategy calculation from CrysAlisPro (Rigaku OD, 2023). Data integration and reduction were also performed using CrysAlisPro, and a numerical absorption correction was applied based on multi-scan or Gaussian methods. The structures were solved by intrinsic phasing using the SHELXT software package and were refined with SHELX.

Powder X-ray diffraction patterns were obtained using a Rigaku Miniflex-II instrument operating with Cu$-K\alpha$ radiation with $\lambda=1.5406\,$\AA $(K\alpha_1)$ and $1.5443\,$\AA $(K\alpha_2)$ at 30 kV and 15 mA. Samples were removed from the glovebox, quickly ground, and attached to a sample holder using Dow-Corning high-vacuum grease. The patterns were collected in air, and the total measurement time was around one hour for each pattern. As expected for air- and/or moisture-sensitive samples, the grinding process accelerates the crystal degradation. In particular, we found that the impurity phase fractions increase with the time that the ground powder is exposed to air. The latter is evident for Pb, see Fig. S4 in the SI, suggesting that this impurity phase may arise from crystal decomposition, instead of being initially incorporated or attached to the single crystal surfaces.

\subsection{Physical property measurements}

Temperature-dependent resistance measurements were performed between 2 and 300\,K with either a Quantum Design Physical Properties Measurement System (PPMS) or a Janis SHI-950 closed-cycle cryostat. In both cases, the AC resistance was measured using LakeShore AC resistance bridges (models 370 and 372), with a frequency of 16.2 Hz and a 3.16 mA excitation current. The temperature was measured by a calibrated Cernox 1030 sensor connected to a LakeShore 336 controller. As the crystals naturally grew with a rod-like morphology, we measured the resistances along the rod axis, i.e., the crystallographic $c$-axis. Samples were quickly removed from the glovebox, and contacts were made by spot welding a 25 $\mu$m thick annealed Pt wire onto the 100 (or equivalently, 010) faces of the crystals in the standard four-point geometry. After spot welding, a small amount of silver epoxy was painted onto the contacts to ensure good mechanical strength, yielding typical contact resistances of $\approx1\,\Omega$.

\subsection{Density functional theory calculations}

Density functional theory \cite{hohenberg1964, kohn1965} (DFT) calculations were performed with the PBE \cite{perdew1996} exchange-correlation functional using a plane-wave basis set and projector augmented wave method \cite{blochl1994}, as implemented in the Vienna Ab-initio Simulation Package \cite{kresse1996, kresse1996e} (VASP). To calculate the total energy and band structure, we used a kinetic energy cutoff of 500 eV, a $\Gamma$-centered Monkhorst-Pack \cite{monkhorst1976} (6$\times$6$\times$7) $k$-mesh, and a Gaussian smearing of 0.05 eV. The ionic positions and unit cell vectors were fully relaxed until the remaining absolute force on each atom was less than 0.01 eV/\AA. For accurate calculation of density of states (DOS), we use a dense (12$\times$12$\times$14) $k$-mesh with the tetrahedron method \cite{blochl1994i}. To calculate the susceptibility functions efficiently \cite{wang2024}, maximally localized Wannier functions (MLWF) \cite{marzari1997, souza2001} and the tight-binding model were constructed with La \textit{sdf}, Pb \textit{p}, and O \textit{p} orbitals to reproduce closely the band structure around the Fermi energy ($E_F$) within $E_F\,\,\pm$ 1 eV. The bare susceptibility function was calculated with eight bands, four valence and four conduction bands, around the Fermi level on a dense (50$\times$50$\times$50) \textit{k}-mesh using the MLWFs. Electron-phonon coupling (EPC) interaction was calculated with density functional perturbative theory \cite{baroni2001} (DFPT) as implemented in Quantum Espresso \cite{giannozzi2009} (QE) using ultra-soft pseudopotentials with a kinetic energy cutoff of 50 Ry and a (2$\times$2$\times$2) $q$-mesh. 

\begin{figure}[ht]
    \centering
    \includegraphics[scale=.9]{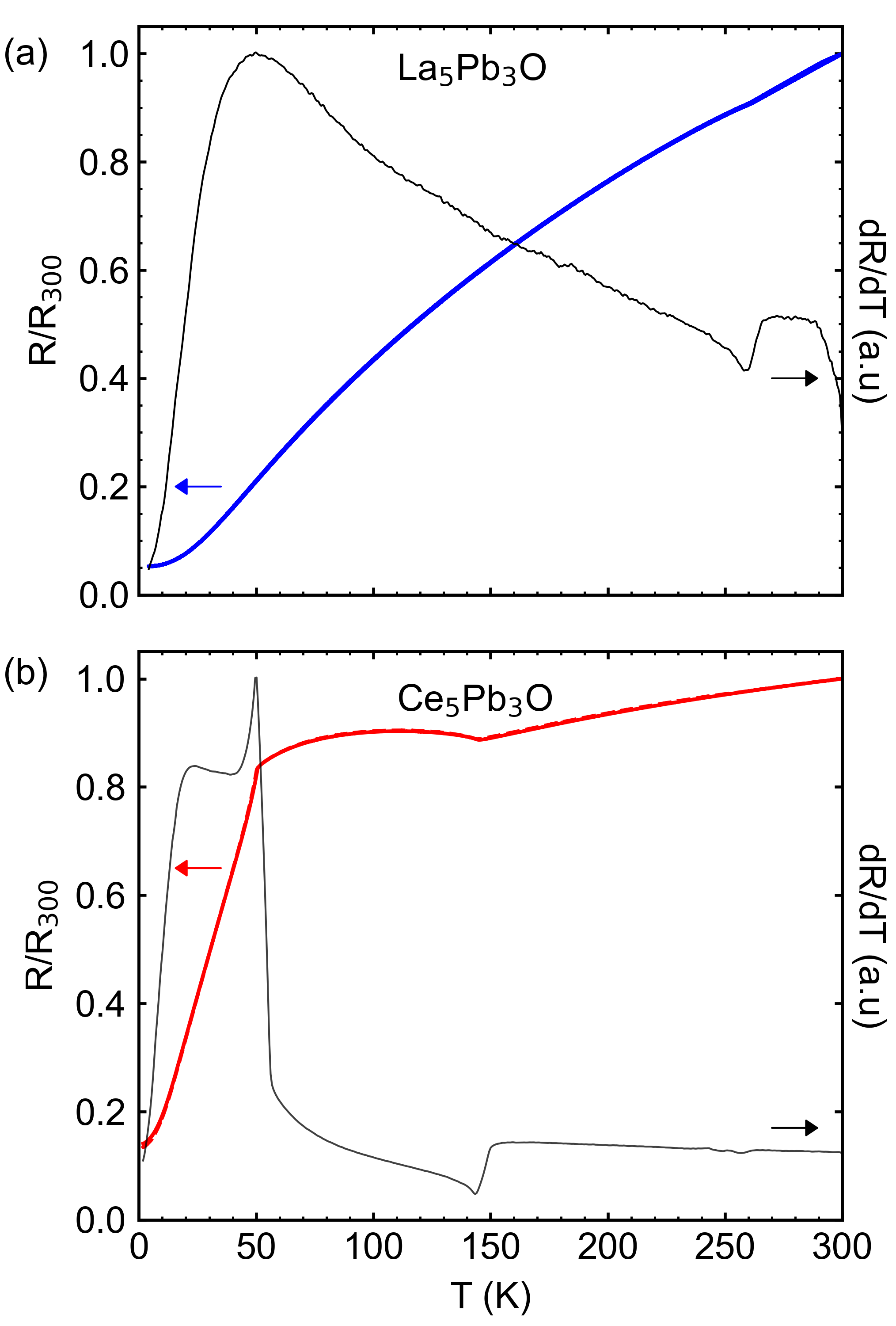}
    \caption{Temperature dependent resistance normalized to its value at $300\,$K (left axis), $R/R_{300}$, for (a) La$_{5}$Pb$_{3}$O and (b) Ce$_{5}$Pb$_{3}$O samples. The data was collected upon cooling and warming, as indicated by dashed and solid lines, respectively, and almost perfectly overlaps in the whole temperature range. Resistance derivatives (right axis), $dR/dT$, are also shown. The downturn in $dR/dT$ that occurs near 300 K in (a) is from imperfect temperature stability near room temperature.}
    \label{fig:resistance}
\end{figure}

\begin{figure*}[ht]
    \centering
    \includegraphics[scale=1.1]{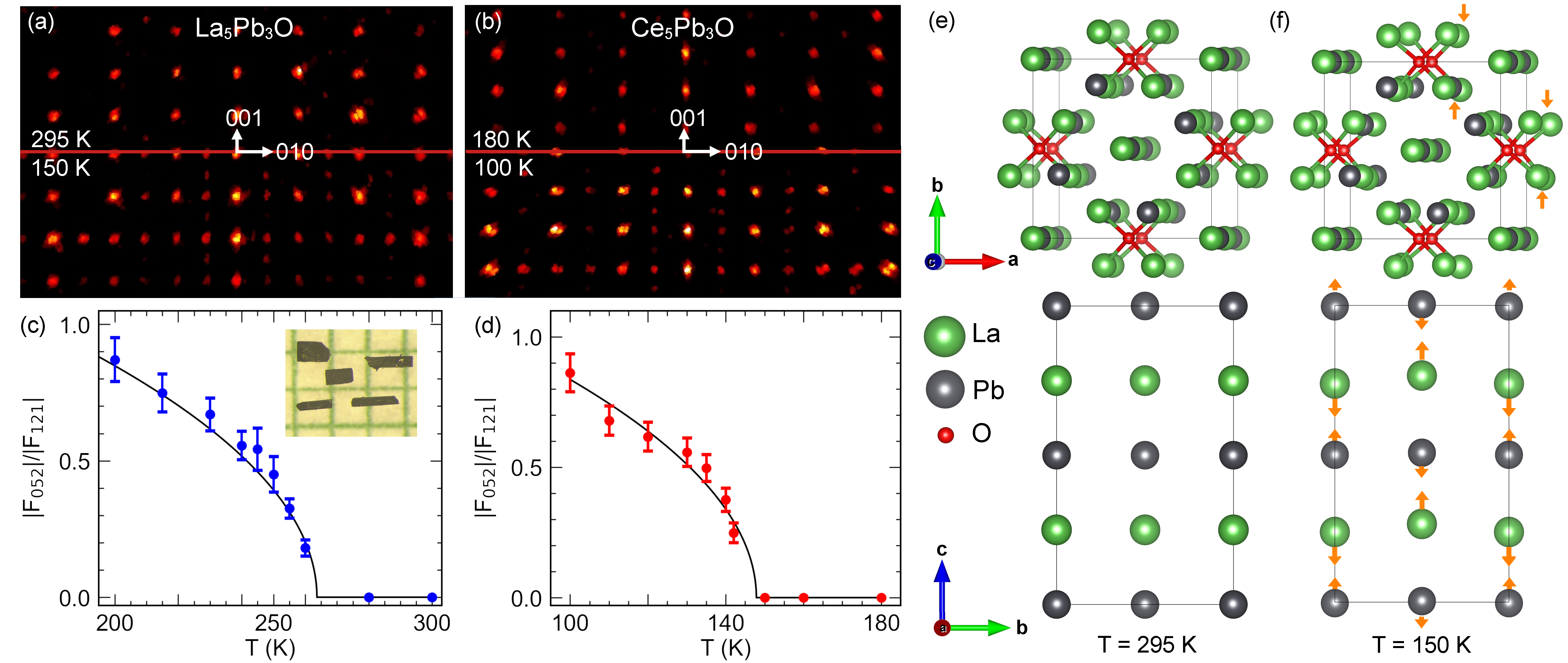}
    \caption{Single crystal diffraction of (a) La$_{5}$Pb$_{3}$O and (b) Ce$_{5}$Pb$_{3}$O samples along the $100$ direction above and below $T_{CDW}$. The additional peaks in the low-temperature phase exhibit odd $k$ and $l = 2n$, consistent with a symmetry reduction from t$I$ to t$P$. Integrated intensity of the (052) reflection normalized to the (121) Bragg peak can serve as a proxy for the order parameter of the phase transition for (c) La$_{5}$Pb$_{3}$O and (d) Ce$_{5}$Pb$_{3}$O. The black lines show power law fits to the data according to the expression $a(T-T_{c})^{-1/2}$. The inset in (c) shows a picture of typical R$_{5}$Pb$_{3}$O single crystals on a mm grid. (e) High- and (f) low-temperature tetragonal structures of La$_{5}$Pb$_{3}$O along $c$- and $a$ crystallographic directions, shown in top and bottom panels, emphasizing the linear La/Pb chains intercalated with O-centered La$_4$ tetrahedra and interpenetrating Pb atoms. To highlight the R-Pb pairs, only R/Pb chains are shown in the bottom panel. The orange arrows lines indicate the main distortions observed in the low-temperature structures of both R$_5$Pb$_3$O compounds.}
    \label{fig:diffraction}
\end{figure*}

\section{Results and discussion}

Fig. \ref{fig:resistance} presents the temperature-dependent resistance of the R$_5$Pb$_3$O single crystals, revealing a phase transition around $\approx $ 260 K and 145 K for R = La and Ce, respectively. At the transition, the resistance shows a clear anomaly, which is modest for La$_5$Pb$_3$O (though clearly observed in the resistance derivative, d\textit{R}/d\textit{T}). In Ce$_5$Pb$_3$O, the signature of the transition is much stronger, as the resistance increases by nearly 5$\%$ between 145\,K and 100\,K. The behavior of the resistance near the transition temperature suggests a partial gapping of electronic bands near the Fermi level, indicating that the phase transition lowers the crystallographic symmetry on cooling, and the behavior in the vicinity of the transition is consistent with CDW ordering \cite{naito1982electrical,song2003charge,PhysRevB.99.235138,ru2008magnetic}. The heating and cooling curves in Fig. \ref{fig:resistance} present no significant thermal hysteresis, in agreement with a second-order phase transition. The observed features are reproducible across multiple samples and were not reported in earlier works on La$_5$Pb$_3$O and Ce$_5$Pb$_3$O, in which transport measurements were not performed \cite{guloy1992, macaluso2004}. 

For Ce$_5$Pb$_3$O, the resistance also shows a sharp decrease upon cooling below 50\,K, which is a typical signature of magnetic ordering and agrees reasonably with the ferrimagnetic transition at $T_c = 46\,$K observed by Macaluso et al. in magnetization measurements \cite{macaluso2004}. We note that transport measurements on our samples that were grown in Al$_2$O$_3$ crucibles, following the method outlined by Macaluso et al., show the magnetic transition at 46 K and also a moderate suppression of the CDW feature near 145 K, see Fig. \ref{fig:EDSRT}. The shifted transitions in samples grown within Al$_2$O$_3$ crucibles can likely be explained by a small amount of Al being incorporated into the crystals (see discussion in the SI) and implies that both the density wave and magnetic ordering transitions may be sensitive to even small chemical substitutions. This topic will be explored in more detail in future work.

\begin{table*}[htbp] %add [H] placement to break table across pages
\begin{ruledtabular}
\caption{\footnotesize{Atomic coordinates and anisotropic displacement parameters ($10^{-2}$\,\AA$^2$) of La$_5$Pb$_3$O and Ce$_5$Pb$_3$O single crystals above and below the CDW transition. In the high-temperature phase, these compounds adopt a body-centered tetragonal structure (space group I$4/mcm$). Below $T_{CDW}$, the structural distortion along the (001) direction results in primitive symmetry, with space group P$4/ncc$. Uncertainties are given in parentheses. }}
\label{tab:structures}
\resizebox{\linewidth}{!}{\begin{tabular}{cclllll|lllll}
\multicolumn{12}{c}{La$_5$Pb$_3$O}  \\ \hline
     &         & \multicolumn{5}{c}{T = 295 K (I$4/mcm$)}       & \multicolumn{5}{c}{T = 150 K (P$4/ncc$)}    \\ \hline
Atom & Wyckoff  & $x$         & $y$        & $z$        & $U_{11}=U_{22}$  & $U_{33}$  & $x$        & $y$        & $z$        & $U_{11}$      & $U_{33}$ \\
La1  & $4c$     & 0           & 0          & 0          & 0.67(1)          & 4.36(4)   & 0          & 0          & 0.01361(5)  & 0.30(1)      & 1.84(3)    \\
La2  & $16l$    & 0.65223(3)  & 0.15223(3) & 0.15024(2) & 1.086(9)         & 0.94(1)   & 0.35357(3) & 0.15810(3) & 0.15022(2)  & 0.50(1)      & 0.59(1)   \\
Pb1  & $4a$     & 0           & 0          & 1/4        & 1.02(1)          & 1.41(2)   & 0          & 0          & 0.25344(2)  & 0.44(9)      & 0.83(1)    \\
Pb2  & $8h$     & 0.35559(22) & 0.14441(2) & 0          & 0.777(8)         & 1.03(1)   & 0.14435(2) & 0.35565(2) & 0           & 0.329(8)     & 0.65(1)     \\
O    & $4b$     & 1/2         & 0          & 1/4        & 1.3(2)           & 1.2(3)    & 1/2        & 0          & 1/4         & 0.5(2)       & 1.7(4)       \\ \hline
\multicolumn{12}{c}{Ce$_5$Pb$_3$O}      \\ \hline
     &         & \multicolumn{5}{c}{T = 180 K (I$4/mcm$)}           & \multicolumn{5}{c}{T = 100 K (P$4/ncc$)}   \\ \hline
Atom &  Wyckoff & $x$        & $y$        & $z$        & $U_{11}=U_{22}$ & $U_{33}$  & $x$        & $y$        & $z$        & $U_{11}$  & $U_{33}$ \\
Ce1  & $4c$     & 0          & 0          & 0          & 0.76(2)         & 2.82(5)   & 0          & 0          & 0.00945(5) & 0.72(2)  & 1.84(3)                                        \\
Ce2  & $16l$    & 0.65301(4)  & 0.15301(4)  & 0.15071(4) & 0.98(2)        & 0.94(2)   & 0.35076(4) & 0.15715(4) & 0.15073(2) & 0.85(1)      & 0.93(2)                                        \\
Pb1 & $4a$      & 0          & 0          & 1/4        & 0.94(2)        & 1.16(2)   & 0          & 0          & 0.25248(2) & 0.80(1)      & 1.07(2)                                        \\
Pb2  & $8h$     & 0.35635(4) & 0.14365(4) & 0          & 0.80(1)        & 0.96(2)   & 0.14351(3) & 0.35649(3) & 0          & 0.72(1)      & 0.94(2)                                       \\
O    & $4b$     & 1/2        & 0          & 1/4        & 1.1(3)          & 1.4(5)      & 1/2        & 0          & 1/4        & 1.1(2)      & 1.7(4)                                        
\end{tabular}}
\end{ruledtabular}
\end{table*}

% previous figure named:  Figures/FiguresTaGrowth/Refinement.png

\begin{figure*}[htbp]
    \centering
    \includegraphics{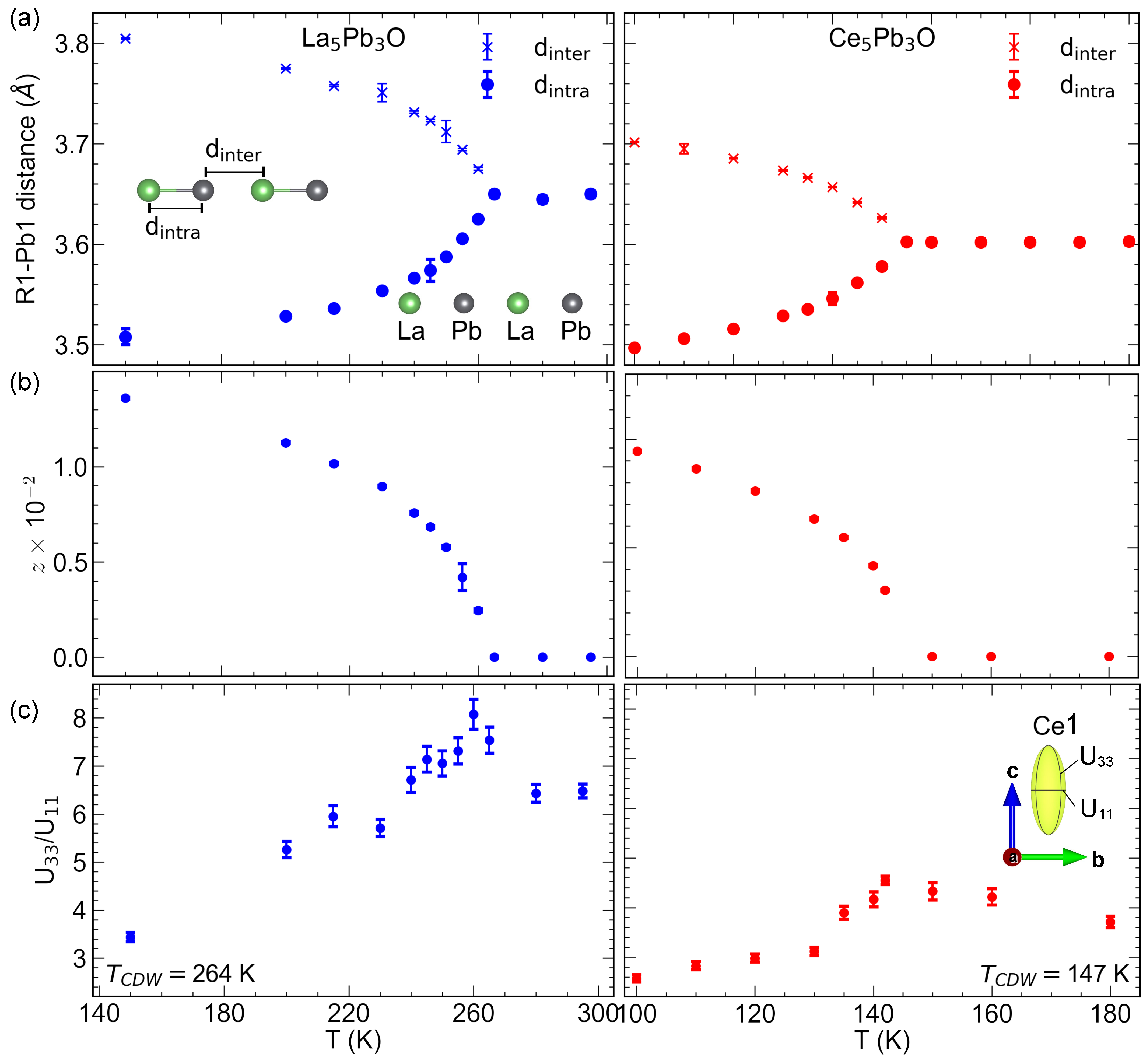}
    \caption{Temperature dependence of the structural parameters for La$_5$Pb$_3$O (left panel) and Ce$_5$Pb$_3$O (right panel). (a) R1-Pb1 distances in the chains along the (001) direction, revealing the formation of R1-Pb1 pairs (i.e., nonequivalent R1-Pb distances within the chains) when cooling through the transition. Intra- and inter-pair distances are represented by circles ($d_{inter}$) and ``x'' ($d_{intra}$) symbols, respectively. (b) Fractional coordinate $z$ of La (left, blue) and Ce (right, red) at the $4c$ sites. Above $T_{CDW}$, this site is symmetry-protected with $z = 0$. (c) The ratio between the principal axes of the thermal ellipsoids along $c$ and $a$ or $b$ directions, $U_{33}$ and $U_{11} = U_{22}$, respectively, exhibiting a maximum at $T_{CDW}$. The transition temperatures shown at the bottom of (c) were estimated from the order parameter fits shown in Figs. \ref{fig:diffraction}(c, d).}
    \label{fig:refinement}
\end{figure*}

Given that this phase transition is observed in both La$_5$Pb$_3$O and Ce$_5$Pb$_3$O (i.e., with both moment-bearing and non-moment-bearing rare-earth elements), it is most likely structural in nature. To explore this further, we measured temperature-dependent single crystal X-ray diffraction from 295\,K to 100\,K. Figures \ref{fig:diffraction}(a, b) show the diffracted intensities in the ($0kl$) plane above and below the transition for R = La and Ce based phases, respectively. The high-temperature diffraction data is consistent with the reported body-centered tetragonal structure (t$I$) of these compounds, space group $I4/mcm$ (n. 140), in which Bragg peaks with $k = 2n +1$ and $l= 2n$ are forbidden. Spots with odd $k$ appear in the low-temperature data, suggesting a transition to a primitive structure (t$P$). Figures \ref{fig:diffraction}(c, d) show the intensity of the (052) reflection, which is forbidden in the high-temperature $I4/mcm$ structure, normalized to the (121) peak intensity. In both compounds, $|F_{\text{052}}|$/$|F_{\text{121}}|$ grows like an order-parameter below the transition temperature, agreeing well with a mean-field (\textit{T}-\textit{T}$_{CDW}$)$^{-1/2}$ temperature dependence, and based on these fits, we estimate $T_{CDW}\approx$ 264\,K and 147\,K for R = La and Ce, respectively. 

The high-temperature structure can be viewed as a network of chains running along the (001) direction, one composed of alternating R/Pb atoms and the other of O-centered R$_4$ tetrahedra with interpenetrating Pb sites, as shown in Fig. \ref{fig:diffraction}(e). At room temperature, R-Pb distances in the linear chain are 3.6503(1) \AA\, and 3.6037 (2) \AA\, for R = La and Ce, respectively. Our diffraction data indicated that, when cooling through the transition, the R atoms are displaced towards the Pb atoms, with R atoms shifted in opposite directions in adjacent chains, as shown in Fig.\ref{fig:diffraction}(f). The displacement of R/Pb atoms results in nonequivalent R-Pb bond distances within each chain, yielding R-Pb pairs. The symmetry breaking at the transition is associated with a change from a body centered to a primitive structure with space group $P4/ncc$ (n. 130), as seen in Fig. \ref{fig:diffraction}(f). In addition to the distortion of R atoms in the (001) direction, the low-temperature structure also features a small in-plane distortion, with the La$_4$O tetrahedra rotated in the $ab$-plane along $c$-axis, as highlighted by orange arrows in the top panel of Fig. \ref{fig:diffraction}(f). The low-temperature structure can be described by a displacement mode that transforms as the irreducible representation $M3-$ of $I4/mcm$ (notation from ISODISTORT) with wave vector $q=(0\,\,0\,\,1)$ \cite{campbell2006, stokes}. Within the R-Pb chains, the intra-pair R-Pb distances are $\sim$ 3.5\,\AA\, for both compounds, while inter-pair distances are $\sim$ 3.8\,\AA\,, and 3.7\,\AA\, for R = La and Ce, respectively. Table \ref{tab:structures} lists the refined atomic positions and displacement parameters for both compounds. Further refinement information is given in Tables \ref{tab:refinementLa5Pb3O} and \ref{tab:refinementCe5Pb3O} in the SI.  

\begin{figure*}[htbp]
    \centering
    \includegraphics[scale=.8]{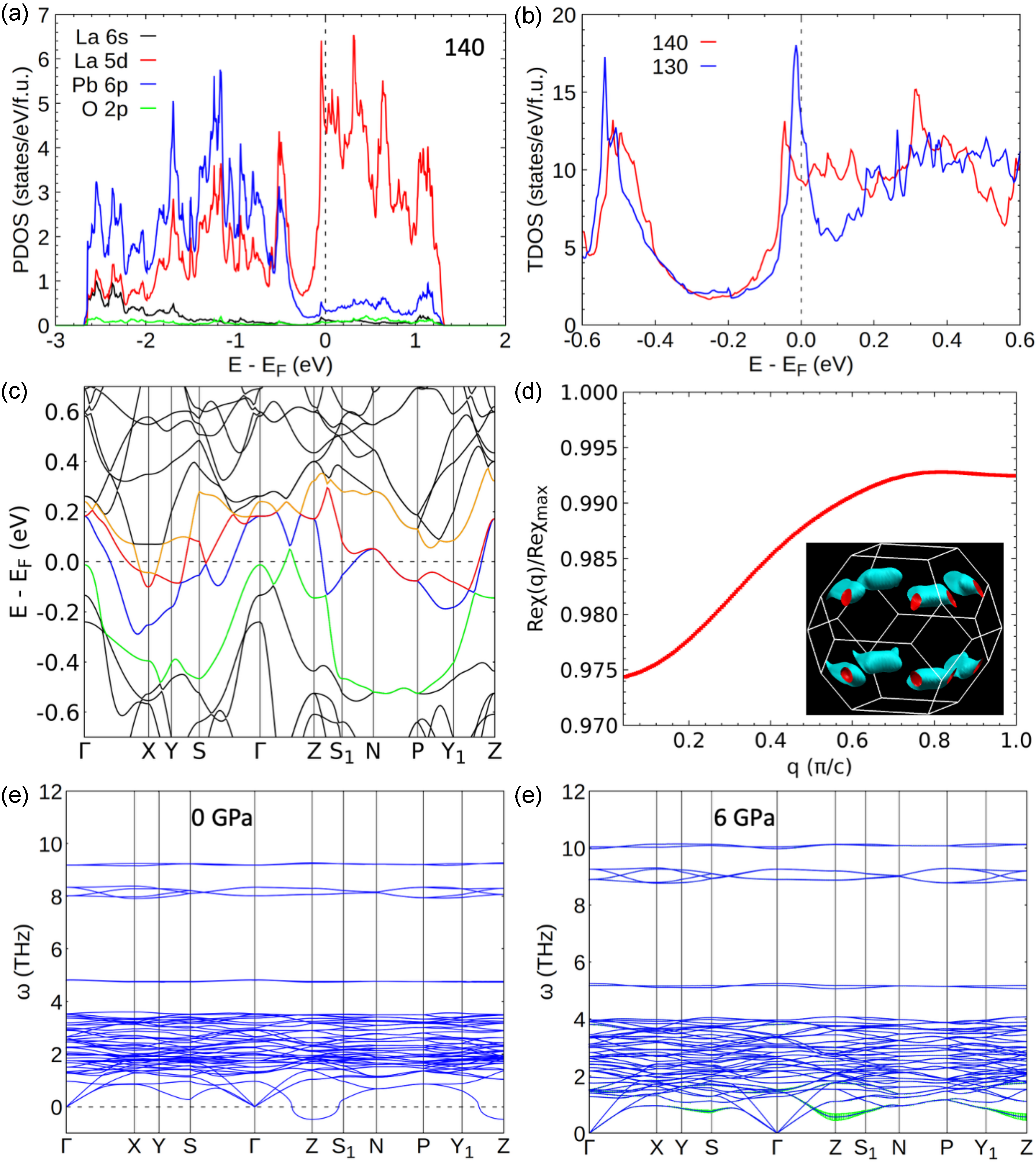}
    \caption{(a) Orbital-projected electronic density of states (PDOS) for the high-temperature tetragonal structure ($tI$, space group 140). (b) Comparison of the total density of states (TDOS) between high- and low-temperature structures, space groups 140 and 130, respectively, around the Fermi energy. (c) Electronic band structure of the high-temperature phase. The four bands crossing $E_F$ are highlighted in different colors. (d) Real part of the bare susceptibility function $\text{Re}\,\chi(\mathbf{q})/\text{Re}\,\chi_{max}$ for the high-temperature structure along the $k_z$ direction. The inset shows the 3D $\text{Re}\,\chi(\mathbf{q})/\text{Re}\,\chi_{max} = 0.999$ at its maximum near the $P$-point. Phonon band structure of the high-temperature structure (e) at equilibrium with 0 GPa pressure and (f) at 6 GPa. The green shade in (f) stands for the magnitude of mode-resolved electron-phonon coupling strength ($\lambda_{q\omega}$).}
    \label{fig:dft}
\end{figure*}

The refinements of our X-ray data show that, in both the low- and high-temperature phases, the R atoms in the linear R-Pb chains (at the $4c$ Wickoff position), namely La1 and Ce1 in Table \ref{tab:structures}, exhibit extremely ellipsoidal thermal displacement parameters, with room temperature values of $U_{33}/U_{11}$ $\approx$ 8 for La$_5$Pb$_3$O and 3.3 for Ce$_5$Pb$_3$O. Such large $U_{33}/U_{11}$ ratios were also observed by Guloy et al. in their initial report outlining the discovery of La$_5$Pb$_3$O \cite{guloy1992}. At that time, the authors attributed the anomalous $U_{33}$ to a disordered La1 site, which they modeled as the La atom occupying a position slightly above or below $z=0$ with approximately 50 $\%$ occupancy at either site. The split site results in La-Pb distances of 3.4690(8)\,\AA\, in the chain, comparable with the La-Pb bond length observed in the low-temperature phase. Here, we also attempted to refine our diffraction data considering a split R1 site, and found that whereas this does decrease $U_{33}$, the overall refinement statistics are not significantly improved. In the absence of a phase transition, the structural model with split R1 sites provides a reasonable explanation for the otherwise anomalous thermal displacements; however, given that the R1 atoms are strongly displaced along the (001) direction by the structural transition, we interpret the large $U_{33}/U_{11}$ values not as static disorder but instead as an indication that the R$_5$Pb$_3$O compounds are close to a structural instability. Therefore, in the following, we consider a structural model for the high-temperature phase with a fully occupied site at the position $4c$, i.e., no split R site.

The refinement of the temperature-dependent diffraction data shows that the R-Pb distance within the linear chain is nearly constant above the transition, as shown in Fig. \ref{fig:refinement}(a). Below $T_{CDW}$, the R1 and Pb1 atoms form pairs, with the R1-Pb1 bond length decreasing with further cooling, while the inter-pair distance increases. As shown in Fig. \ref{fig:refinement}(a), the R-Pb distance behaves as an order parameter for the phase transition, like the intensity of the $(0,k=2n+1, l=2n)$ peaks shown in Figs. \ref{fig:diffraction}(c, d). The distortion of the R atoms in the low-temperature phase is easily seen through their fractional coordinates $z$, which increases on cooling below the transition, see Fig. \ref{fig:refinement}(b). Especially interesting, we find that the thermal displacement parameters of the La1 and Ce1 atoms are most ellipsoidal (i.e., $U_{33}/U_{11}$ is maximized) at the transition temperature, as shown in Fig. \ref{fig:refinement}(c). These three features, i) the increasingly asymmetric R-Pb distances within each chain -- Fig. \ref{fig:refinement}(a), ii) the systematic shift in the R1 fractional $z$ coordinate below the transition -- Fig. \ref{fig:refinement}(b), and iii) the maximum in $U_{33}/U_{11}$ at the transition temperature, all suggest that in the high-temperature phase the R1 atoms sit at the (0, 0, 0) position, with equal bond lengths to the adjacent Pb1 atoms, but exhibit highly anisotropic vibrations along the \textit{c}-axis. The temperature dependence of $U_{33}/U_{11}$ shows that the amplitude of the ellipsoidal R1 vibrations in the high-temperature structure increases on cooling and is maximized at the critical temperature. Below $T_{CDW}$, $U_{33}/U_{11}$ decreases on cooling. For instance, at 150\,K, $U_{33}/U_{11}$ is less than half of its maximum value for La$_5$Pb$_3$O. Given that the structural change is associated with a distortion of the R1 atoms along the \textit{z} direction, such behavior is consistent with the softening of a phonon mode as the transition's critical temperature is approached, and is in line with the expectation for a second-order CDW-like phase transition. 
 
Our diffraction data support that the R$_5$Pb$_3$O series hosts a CDW below 260\,K and 145\,K for R = La and Ce, respectively, with the symmetry reduction from t$I$ to t$P$ occurring as the 1D R/Pb chains form R-Pb pairs. The structural transition is accompanied by the gapping of some bands near the Fermi surface, corresponding to the upturn evident in the resistance data -- Figs. \ref{fig:resistance}(a, b). To further support our experimental findings, we calculated the electron, phonon, and electron-phonon coupling properties from first-principles methods. 

Fig. \ref{fig:dft}(a) presents the orbital projected electronic density of states (PDOS) for the high-temperature structure (space group 140). The Pb $6p$ and La $5d$ orbitals hybridize strongly below the Fermi level ($E_F$) from -2.7 to -0.5 eV, while La $5d$ orbitals dominate the states just below and above $E_F$ from -0.5 to +1.3 eV. A sizable pseudogap is formed around -0.2 eV, whose anti-bonding states are gradually filled towards $E_F$. From the electronic band structure in Fig. \ref{fig:dft}(c), four bands (colored in green, blue, red, and orange) cross $E_F$. The separation between the lower and upper bands around -0.2 eV corresponds to the pseudogap in the DOS, and the filling of the anti-bonding states at $E_F$ hints at a potential instability. Indeed, the calculated phonon band structure in Fig. \ref{fig:dft}(e) shows an imaginary mode at the $Z$ point. By following the eigenvector of this imaginary phonon in the supercell for ionic relaxation, we found that a primitive tetragonal structure ($tP$, space group 130) is 10.8 meV per formula unit more stable than the high-temperature body-centered tetragonal structure ($tI$, space group 140). The doubling of the periodicity along the $c$-axis from $tI$ to $tP$ structures corresponds to the imaginary mode at $Z$. The calculated structural distortion and corresponding equilibrium structure agree well with the low-temperature phase experimentally observed. 

Fig. \ref{fig:dft}(b) shows the total density of states (TDOS) around the pseudogap region. By adopting the low-temperature structure ($tP$, space group 130), the size of the pseudogap at -0.2 eV increases as the bonding/anti-bonding states are respectively pushed to lower/higher energies, providing greater electronic stability. Also, in the low-temperature phase, a local minimum is formed just above $E_F$ at +0.1 eV. As a consequence of this reduction in the DOS near 0.1 eV, initially unoccupied bonding states are lowered towards $E_F$, also bringing extra electronic stability. The higher DOS at $E_F$ in the low-temperature structure, compared to the high-temperature phase, is a consequence of the changes above and below $E_F$, i.e., the net effect of both the widening of the lower-energy pseudogap and the formation of the new higher-energy local minimum.

To understand the origin of the CDW in the R$_5$Pb$_3$O series, we investigated both Fermi surface nesting (FSN) and electron-phonon coupling (EPC) of La$_5$Pb$_3$O. From the electronic band structure of the high-temperature phase in Fig. \ref{fig:dft}(c), we constructed Wannier functions and calculated the bare susceptibility function $\chi(\mathbf{q})$ on a dense $k$-point mesh \cite{wang2024}. The 1D $\text{Re}\,\chi(\mathbf{q})/\text{Re}\,\chi_{max}$ along the $k_z$ direction, chosen because the primary distortion is along the (001) direction and displayed in Fig. \ref{fig:dft}(d), shows only a broad peak away from the $Z$ point, and the 3D $\text{Re}\,\chi(\mathbf{q})/\text{Re}\,\chi_{max} = 0.999$, see the inset of Fig. \ref{fig:dft}(d), shows that $\text{Re}\,\chi(\mathbf{q})$ is maximized near the $P$ point instead of $Z$. Based on these calculations, the CDW in La$_5$Pb$_3$O is unlikely to be driven by Fermi surface nesting. 

As shown in Fig. \ref{fig:dft}(f), the calculations indicated that a (theoretical) pressure of 6 GPa stabilizes the imaginary phonon mode at the $Z$ point. The mode-resolved electron-phonon coupling (EPC) $\lambda_{q\omega}$ in the compressed structure, indicated by the green shade in Fig. \ref{fig:dft}(f), clearly shows that the EPC is the strongest for the lowest optical phonon mode at the $Z$ point, followed by the phonons at $S$. When pressure is removed, the strong EPC softens the phonon at the $Z$ point enough to make it imaginary and to drive the CDW. Therefore, these calculations indicate that the CDW in the R$_5$Pb$_3$O series is driven by EPC, instead of Fermi surface nesting \cite{wang2024}, similar to what is observed in NbSe$_2$ \cite{johannes2006, zhu2015}. The prediction that pressure stabilizes the imaginary mode is also qualitatively in line with the experimental observation that the transition in Ce$_5$Pb$_3$O occurs $\approx$ 100 K below that in La$_5$Pb$_3$O, which might be attributed to chemical pressure associated with the lanthanide contraction.

\section{Summary}

R$_5$Pb$_3$O compounds were discovered decades ago and have remained mostly unexplored to date. At room temperature, these compounds adopt a tetragonal structure featuring linear R-Pb chains along the $c$-axis. Transport measurements reveal a second-order phase transition at 260\,K and 145\,K for R = La and Ce, respectively. Temperature-dependent single crystal X-ray diffraction revealed that the phase transition is structural in nature, and that the primary distortion is a modulation within the R-Pb chains. The anisotropic displacement parameters of the R atoms in the chain are highly ellipsoidal and are maximized at the transition temperature, which we interpret as a signature of the phonon softening associated with the structural transition. Density functional theory calculations for the high-temperature structure of La$_5$Pb$_3$O reveal an imaginary phonon mode at the \textit{Z}-point, consistent with the experimentally observed structural instability. Altogether, our experimental data and theoretical calculations support that the R$_5$Pb$_3$O (R = La and Ce) series undergo a charge density wave (CDW) ordering that is driven by electron-phonon coupling. Both experimental and theoretical results suggest that the CDW transition temperature may be strongly suppressed by physical or chemical pressure, possibly making this system an attractive platform for exploring competing orders. It will likewise be worthwhile to explore how the CDW and magnetic transitions evolve with heavier rare-earth members in future work.

\begin{acknowledgments}
Work at Ames National Laboratory was supported by the U.S. Department of Energy (DOE), Basic Energy Sciences, Division of Materials Sciences \& Engineering, under Contract No. AC02-07CH11358. RFSP was financed, in part, by the São Paulo Research Foundation (FAPESP), Brasil. Process Number 2024/08497-6. HAH was supported by the U.S. Department of Energy Office of Science Science Undergraduate Laboratory Internship (SULI) program under its contract with Iowa State University, Contract No. DE-AC02-07CH11358. The computational part of this research used resources of the National Energy Research Scientific Computing Center (NERSC), a DOE Office of Science User Facility.
\end{acknowledgments}

*corresponding authors' email: slade@ameslab.gov, rafaelap@iastate.edu

\bibliography{bib}

%%%%%%%%%%%%%%%%%%%%%%%%%% supplementary material starts here
\renewcommand{\thefigure}{S\arabic{figure}}
\setcounter{figure}{0}

\renewcommand{\thetable}{S\arabic{table}}
\setcounter{table}{0}

\clearpage

\onecolumngrid
\section{Supplementary information}

\subsection{Powder X-ray diffraction and phase indexing}

\begin{figure}[htbp]
    \centering
    \includegraphics{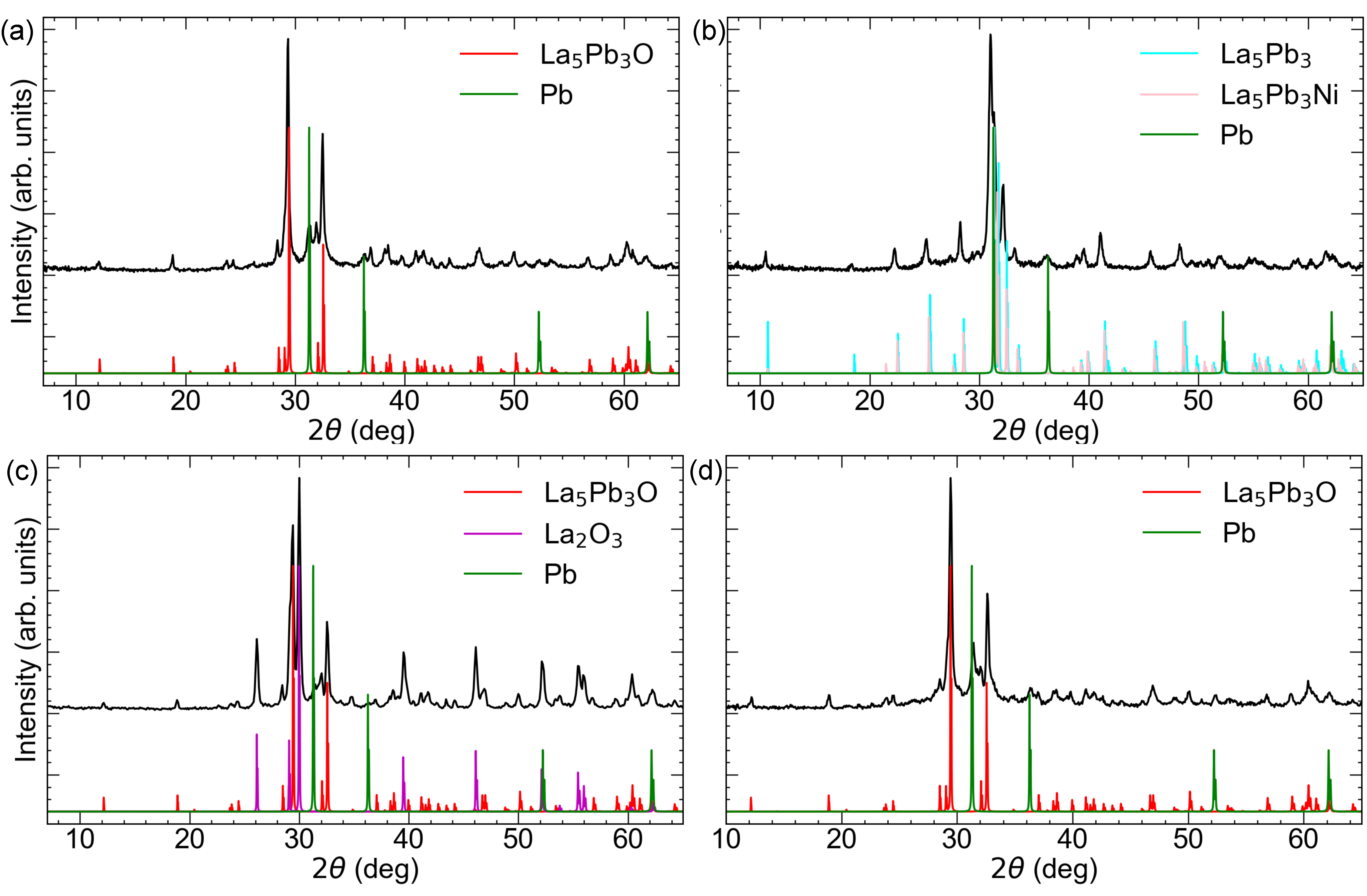}
    \caption{Room temperature powder X-ray diffraction patterns for La$_5$Pb$_3$O samples samples obtained by solution growth with nominal composition (a) La$_{45}$Ni$_{50}$Pb$_5$ (\#1), (b) La$_{45}$Ni$_{50}$Pb$_5$ (\#2), (c) La$_{45}$Ni$_{50}$Pb$_5$ + 3.5\% La$_2$O$_3$, and (d) La$_{45}$Ni$_{50}$Pb$_5$ + 2.5\% PbO in the starting melt. The reactions were contained in Ta crucibles. Here, \#1 and \#2 refer to two separate trials using the protocol outlined in the Experimental Details section. The theoretical patterns of La$_5$Pb$_3$O (high-temperature structure, space group $I4/mcm$), La$_2$O$_3$, and \textit{fcc} Pb, are also shown.}
    \label{fig:powderLaAttempts}
\end{figure}

\begin{figure}[htbp]
    \centering
    \includegraphics{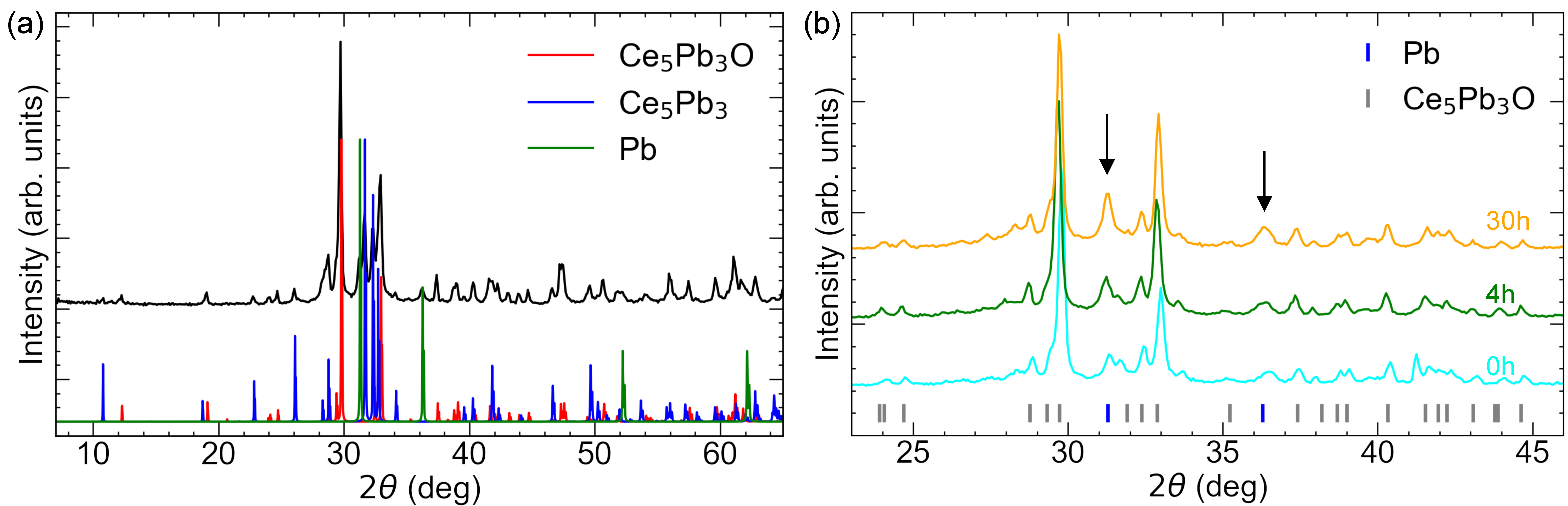}
    \caption{(a) Room temperature powder X-ray diffraction pattern for Ce$_5$Pb$_3$O samples obtained by solution growth in a Ta crucible with nominal composition Ce$_{45}$Co$_{50}$Pb$_5$ + 2\% PbO in the starting melt, as described in the Experimental Details section. The theoretical patterns for Ce$_5$Pb$_3$O (high-temperature phase from ref \cite{macaluso2004}, space group $I4/mcm$), Ce$_5$Pb$_3$, and \textit{fcc} Pb are also shown. (b) Diffraction patterns obtained in ground crystals that were exposed to air for 0, 4 and 30 hours. Expected peak positions of Ce$_5$Pb$_3$O (space group $I4/mcm$) and Pb phases are also shown. Black arrows highlight the Pb peaks, whose intensities increase as crystals degrade in air.}    
    \label{fig:powderZX711}
\end{figure}

Powder X-ray diffraction patterns obtained on the products of the different growths discussed in the \textit{Experimental Details}
section of the main text are presented in Figs. \ref{fig:powderLaAttempts} and \ref{fig:powderZX711}(a). The primary Bragg peaks agree with the body-centered tetragonal structure expected for (room temperature) R$_5$Pb$_3$O; however, a small fraction, 7 to 10 $\%$, of Pb impurity was found in all growth attempts. Such a quantity of Pb would likely be detected in resistance measurements, where we would anticipate to observe a superconducting transition at 7 K corresponding to the $T_{\text{c}}$ of elemental Pb. Yet, our measurements (see Figure 1 of the main text) do not show any evidence for a superconducting transition, suggesting Pb is not present on/within the samples used to measure resistance. Considering this, and given that the R$_5$Pb$_3$O is very air sensitive, we speculate that Pb is a decomposition product of the R$_5$Pb$_3$O and is present in substantial fraction in the PXRD owing to the accelerated decomposition (due to high surface area) associated with powdered samples. In support of this hypothesis, Figure \ref{fig:powderZX711}(b) shows a series of PXRD patterns for a Ce$_5$Pb$_3$O sample that were collected after 0 h, 4h, and 30 h of air exposure, and we indeed find that the fraction of Pb impurity increases with the time that ground crystals are exposed to air. 

\clearpage
\subsection{Comparison of Ce$_5$Pb$_3$O single crystals grown in Al$_2$O$_3$ and Ta crucibles}

As discussed in the main text, we grew R$_5$Pb$_3$O single crystals in reactions contained in Ta crucibles (our own procedure) and also by following the exact procedure outlined by Yan and Macaluso et al. \cite{yan2015, macaluso2004}, in which the reactions were contained in Al$_2$O$_3$ crucibles. In the later, it is likely that the reaction between the rare earth elements and the crucible is responsible for the introduction of oxygen into the melt; however, such reactions will also produce Al metal, which could feasibly be incorporated into the R$_5$Pb$_3$O products. To investigate the possibility of Al contamination in R$_5$Pb$_3$O single crystals grown in Al$_2$O$_3$ crucibles, we performed energy dispersive X-ray spectroscopy (EDS) quantitative elemental analysis using an EDS detector (Thermo NORAN Microanalysis System, model C10001) attached to a JEOL scanning-electron microscope (SEM). An acceleration voltage of 21 kV and a working distance of 10 mm were used for all measurements. A R$_5$Pb$_3$O single crystal grown using a Ta tube was used as a standard for the La/Ce and Pb, while a LaAl$_2$ single crystal was used as a standard for Al. The spectra were fit using NIST-DTSA II microscopium software. 

\begin{figure}[htbp]
    \centering
    \includegraphics[scale=1]{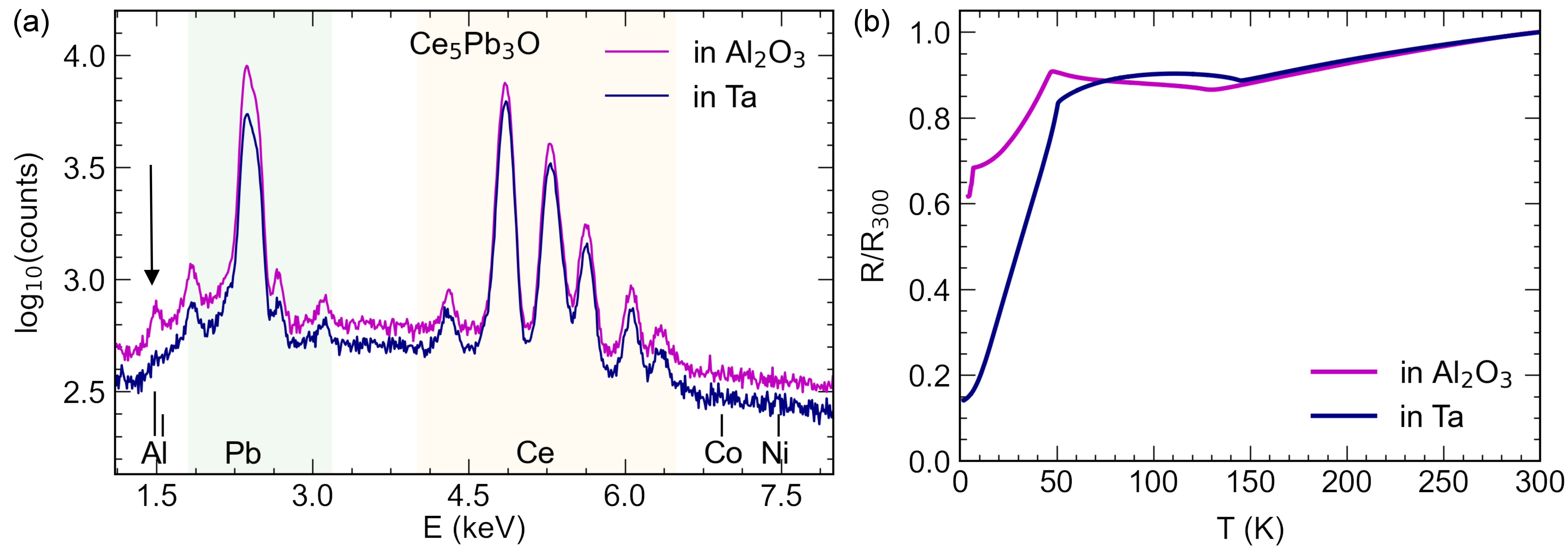}
    \caption{(a) EDS spectra obtained on Ce$_5$Pb$_3$O single crystals grown in Al$_2$O$_3$ and Ta crucibles. Green and orange shaded areas indicate the position of Pb-M and Ce-L spectral lines, respectively. The K-lines of Al, Co, and Ni are also shown. The black arrow highlights the peak associated with Al. (b) Temperature dependent resistance normalized to its value at $300\,$K, $R/R_{300}$, for Ce$_5$Pb$_3$O samples grown using Al$_2$O$_3$ and Ta crucibles.}
    \label{fig:EDSRT}
\end{figure}

As shown in Fig. \ref{fig:EDSRT}(a), the spectra obtained on Ce$_5$Pb$_3$O  samples grown in Al$_2$O$_3$ crucibles exhibit a small, but non-negligible, amount of Al contamination, whereas no Al is detected in the samples grown in Ta crucibles. Based on analysis with standards, we estimate a maximum of 4\% Al contamination. Similar contamination was observed in La$_5$Pb$_3$O crystals grown in Al$_2$O$_3$ crucibles. Single crystal X-ray diffraction data collected on these crystals do not provide substantial evidence for disorder or interstitial atoms, supporting that the degree of Al incorporation is low. Unfortunately, this restricts our ability to directly establish how the Al is incorporated into the structure. Nevertheless, the effect of the Al contamination is easily observed through the physical properties, as outlined below. Finally, we also note that, despite being included in the initial mixtures for both Ta- and Al$_2$O$_3$-grown samples, the EDS spectra show that neither Co nor Ni are detected in the single crystals. 

\begin{figure}[htbp]
    \centering
    \includegraphics[scale=1.2]{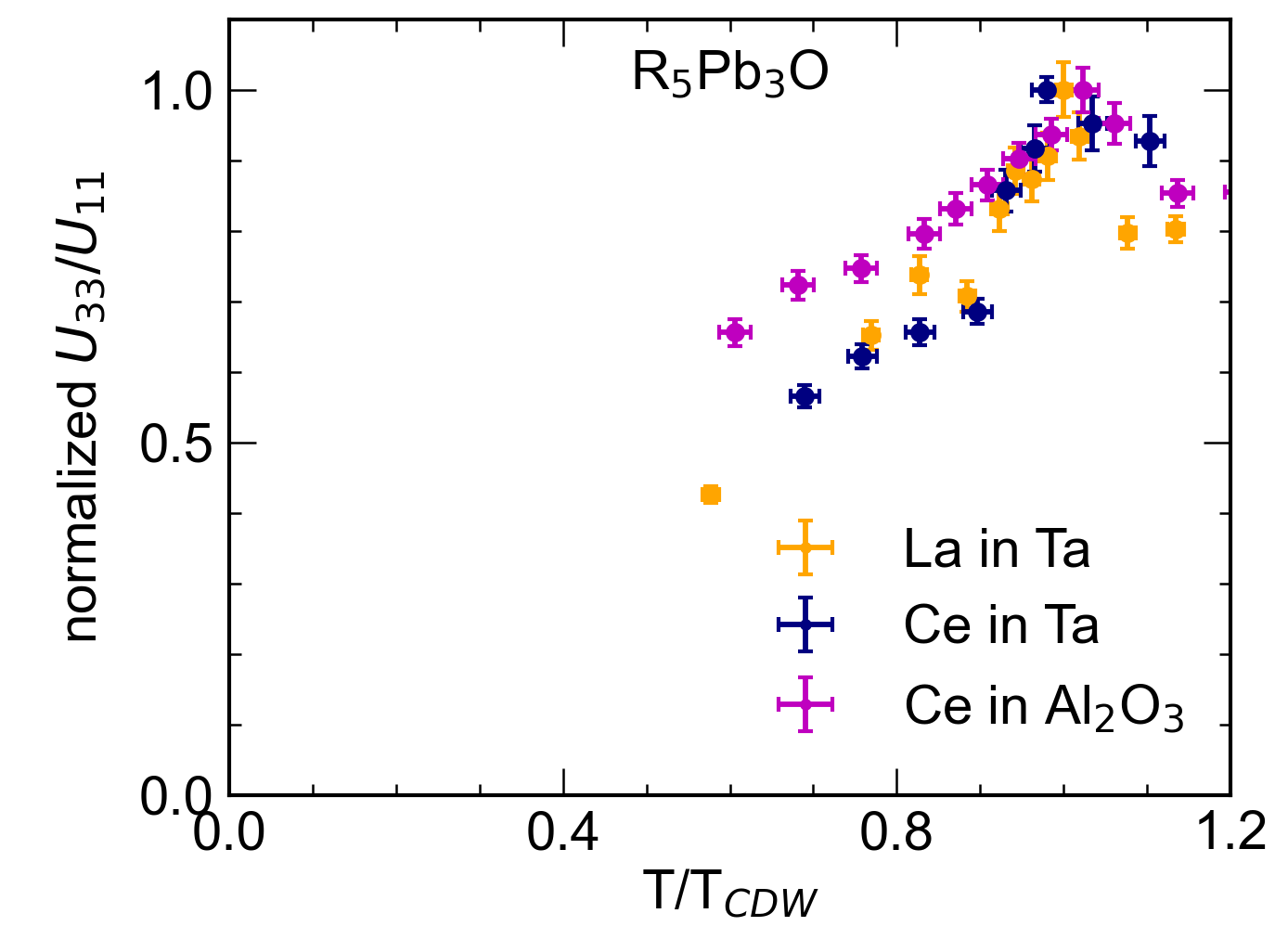}
    \caption{Normalized ratio between the principal axes of the thermal ellipsoids along $c$ and $a$ or $b$ directions, $U_{33}$ and $U_{11} = U_{22}$, respectively, as a function of effective temperature, $T/T_{CDW}$, for R$_{5}$Pb$_3$O samples grown in Ta and Al$_2$O$_3$ crucibles.}
    \label{fig:norm}
\end{figure}

Fig. \ref{fig:EDSRT}(b) compares the temperature-dependent resistance of Ce$_5$Pb$_3$O crystals grown using either Ta or Al$_2$O$_3$. Clear differences are observed between the two samples. The residual-resistance-ratios, RRR = $R(300\,\text{K})/R(4\,\text{K})$, are $\sim$ 7.2 and 1.6 for samples grown in Ta and Al$_2$O$_3$, respectively, where the much smaller value observed for Al-contaminated samples is consistent with disorder. Below $T_{CDW}$, the temperature dependence of the normalized ratio $U_{33}/U_{11}$, shown in Fig. \ref{fig:norm}, is larger for samples grown in Al$_2$O$_3$, whereas both R = La and Ce samples obtained from Ta crucibles exhibit similar behavior. In addition to crystal quality, the Al presence also affects the transition temperatures. The CDW is suppressed by $\sim$ 15 K, while the magnetic transition shifts down by $\sim$ 4 K. That both transition temperatures and RRR are different in the Al$_2$O$_3$-grown samples supports that the Al is chemically incorporated into the sample, as opposed to being a surface or filamentary second phase. In addition to Al contamination, samples grown in Al$_2$O$_3$ also exhibit a small Pb impurity (likely on the surfaces), as indicated by the partial superconducting transition observed in Fig. \ref{fig:EDSRT}(b) at $\approx$ 6.9 K. Interestingly, these results may suggest that both the density wave and magnetic ordering transitions in the R$_5$Pb$_3$O family are sensitive to even small chemical substitution, motivating a future work on controlled doping studies of this family.

\clearpage
\subsection{Crystalographic refinement details}

Tables \ref{tab:refinementLa5Pb3O} and \ref{tab:refinementCe5Pb3O} provide information regarding the structural refinement of single crystal data for R$_5$Pb$_3$O samples.

\begin{table*}[htbp]
    \centering
    \begin{ruledtabular}
    \caption{Single crystal data and structural refinement information for La$_5$Pb$_3$O.}
    \begin{tabular}{lll}
                                            & \multicolumn{1}{c}{295 K}     & \multicolumn{1}{c}{150 K}    \\ \hline
Refined formula                             & La$_5$Pb$_3$O                 & La$_5$Pb$_3$O                  \\
F.W (g/mol)                                 & 1332.12                       & 1332.12                        \\
Space group, $Z$                            & $I4/mcm$, 4                   & $P4/ncc$, 4           \\
$a=b$ (\AA)                                 & 8.7012 (2)                    & 8.7022 (1)                   \\
$c$ (\AA)                                   & 14.6011 (5)                   & 14.6256 (3)                   \\
Volume (\AA$^3$)                            & 1105.46 (6)                   & 1107.58 (4)                   \\
$\theta$ range ($^\circ$)                    & 2.612-31.92                  & 2.612-31.882                   \\
No. reflections, $R_{int}$                  & 14769, 0.0561                  & 30067, 0.0760                        \\
No. independent reflections                 & 1009                           & 1847                          \\
No. parameters                              & 18                            & 24                           \\
$R_1$, $wR_2$ [$I>2d(I)$]                   & 0.0247, 0.0527                 & 0.0311, 0.0775               \\
Goodness of fit                             & 1.256                         & 1.226                       \\
Diffraction peak and role ($e^{-}/\AA^{3}$) & 2.39, -1.92                  & 5.88, -2.88                 
\end{tabular}
\label{tab:refinementLa5Pb3O}
    \end{ruledtabular}
\end{table*}

\begin{table*}[htbp]
    \centering
    \begin{ruledtabular}
    \caption{Single crystal data and structural refinement information for Ce$_5$Pb$_3$O.}
    \begin{tabular}{lll}
                                            & \multicolumn{1}{c}{180 K}     & \multicolumn{1}{c}{100 K}    \\ \hline
Refined formula                             & Ce$_5$Pb$_3$O                 & Ce$_5$Pb$_3$O                  \\
F.W (g/mol)                                 & 1338.17                       & 1338.17                        \\
Space group, $Z$                            & $I4/mcm$, 4                   & $P4/ncc$, 4           \\
$a=b$ (\AA)                                 & 8.5888 (1)                    & 8.5791 (1)                   \\
$c$ (\AA)                                   & 14.3984 (4)                   & 14.3857 (4)                   \\
Volume (\AA$^3$)                            & 1062.12 (4)                   & 1058.80 (4)                   \\
$\theta$ range ($^\circ$)                    & 2.13-32.081                  & 2.159-31.771                   \\
No. reflections, $R_{int}$                  & 15041, 0.0762                  & 28543, 0.0830                        \\
No. independent reflections                 & 956                           & 1751                          \\
No. parameters                              & 18                            & 24                           \\
$R_1$, $wR_2$ [$I>2d(I)$]                   & 0.0286, 0.0896                 & 0.0316, 0.1058               \\
Goodness of fit                             & 1.151                         & 1.312                       \\
Diffraction peak and role ($e^{-}/\AA^{3}$) & 4.33, -3.28                  & 5.18, -2.95                 
\end{tabular}
\label{tab:refinementCe5Pb3O}
    \end{ruledtabular}
\end{table*}

\end{document}